\documentclass[aps,pra,twocolumn,floatfix,10pt]{revtex4-1}
\pdfoutput=1

\usepackage{framed}
\usepackage{graphicx}
\usepackage{amsmath}
\usepackage{epsfig}
\usepackage{helvet}
\usepackage{amssymb}
\usepackage{amsthm}
\usepackage{color}
\usepackage{enumerate}
\usepackage[scriptsize]{subfigure}
\usepackage{times}
\usepackage{dsfont}
\usepackage{bbm}
\usepackage{mathrsfs}

\def\ket#1{\mathinner{\left|{#1}\right\rangle}}
\def\braket#1{\mathinner{\left\langle{#1}\right\rangle}}

\def\bravert{\egroup\,\vrule\,\bgroup}

\newcommand{\dg}{\dagger}

\newcommand{\pd}{\partial}
\newcommand{\id}{\mathds{1}}

\newcommand{\norm}[1]{\left\lVert#1\right\rVert}

\begin{document}

\title{Tachyonic quench in a free bosonic field theory}

\author{Sebasti{\' a}n Montes}
\email{s.montes@csic.es}
\affiliation{Instituto de F\'{i}sica Te\'orica (IFT), UAM-CSIC, Madrid, Spain}

\author{Javier Rodr\'{\i}guez-Laguna}
\affiliation{Dept. of Fundamental Physics, Universidad Nacional de
  Educaci\'on a Distancia (UNED), Madrid, Spain}

\author{Germ\'an Sierra}
\affiliation{Instituto de F\'{i}sica Te\'orica (IFT), UAM-CSIC, Madrid, Spain}

\date{\today}

\begin{abstract} 
We present a characterization of a bosonic field theory driven by a free (Gaussian) tachyonic Hamiltonian. This regime is obtained from a theory describing two coupled bosonic fields after a regular quench. Relevant physical quantities such as simple correlators, entanglement entropies, and the mutual information of disconnected subregions are computed. We show that the causal structure resembles a critical (massless) quench. For short times, physical quantities also resemble critical quenches. However, exponential divergences end up dominating the dynamics in a very characteristic way. This is related to the fact that the low-frequency modes do not equilibrate. Some applications and extensions are outlined. 
\end{abstract}

\maketitle


\section{Introduction}

The study of closed quantum systems out of equilibrium has been developed extensively in the past few decades \cite{Experiments1, CardyCalabrese1, GogolinEisert}, even though the main ideas can be traced back to the origins of quantum theory. This resurgence is due to the development of both experimental and theoretical tools that have sparked new interest in the foundations of statistical mechanics. Among the virtually infinite number of ways of driving a system out of equilibrium, one of the most popular ones is the so-called (global) quench. In this protocol, the system is prepared in the ground state $\ket{0}$ of an initial Hamiltonian $H_0$, which is then let evolve unitarily according to
\begin{equation}
\ket{\psi(t)} = \exp(-itH_1)\ket{0},\quad t\geq 0,
\end{equation}
where $H_1$ usually differs from $H_0$ in one or several of its external parameters (such as magnetic field, pressure, or on-site repulsion). The evolution of different physical properties (entanglement entropy, magnetization, extensive order parameters, etc.) can then be studied to determine the asymptotic dynamics of the system.

On the other hand, tachyonic systems have a much more exotic history. They were original proposed as field theories describing particles that could have a group velocity larger than the speed of light \cite{MetaRel,Feinberg, AronsSudarshan, DharSudarshan}. They are characterized by the dispersion relation
\begin{equation}
E^2 = p^2 - \mu^2,
\end{equation}
where $\mu$ is a real parameter. However, they were later understood on more physical grounds as instabilities that could still preserve Lorentz causality \cite{AKS} and be related instead to symmetry breaking or condensation processes \cite{Peskin, Sen}.

In this paper, we present a characterization of a bosonic field theory driven by a tachyonic Hamiltonian. We will obtain this regime from a theory describing two coupled bosonic fields after a regular quench. We will focus on free (Gaussian) systems \cite{Tegmark, Sotiriadis1, Sotiriadis2}, so that everything can be characterized analytically.  To the best of our knowledge, this type of quench has not been studied in the context of many-body quantum physics. Even though the driving Hamiltonian is unbounded from below, we will argue that the corresponding unitary evolution is well-defined. In particular, we will characterize the evolution of simple correlators, the entanglement entropy of a block, and the mutual information between disconnected subregions. We will show that the causal structure is very similar to the one obtained after a critical quench \cite{Sotiriadis2, LCpropagation, CardyCalabrese}. For short times, physical quantities also resemble the evolution of a massless quench. However, the resulting evolution will be dominated by exponential divergences that prevent the system from being close to a steady state regime. In other words, the system will not equilibrate, even in the sense of generalized Gibbs ensembles \cite{GogolinEisert, GGE}. Some applications and extensions will then be outlined.


\section{Coupled bosonic QFT}

Consider two bosonic fields described by the Lagrangian density
\begin{equation}
\mathcal{L} = \frac{1}{2}(\pd_\mu \Phi_1)^2 + \frac{1}{2}(\pd_\mu \Phi_2)^2 -\frac{m^2}{2}(\Phi_1^2 + \Phi_2^2) - g\Phi_1 \Phi_2.
\label{coupledBosons}
\end{equation}
We can rewrite this Gaussian theory using another set of bosonic fields $\Phi_\pm = (\Phi_1 \pm \Phi_2)/\sqrt{2}$ to obtain
\begin{align}
\mathcal{L} =\frac{1}{2}(\pd_\mu \Phi_+)^2 + \frac{1}{2}(\pd_\mu \Phi_-)^2 - \frac{1}{2}m_+^2\Phi_+^2 - \frac{1}{2}m_-^2\Phi_-^2,
\end{align}
where $m_\pm^2 = m^2 \pm g$. In the new variables, the fields are decoupled and their properties solely determined by their respective masses.

This seemly trivial manipulation allows us to translate the dynamics from the coupling of the original fields to the masses of the new ones. In particular, note that quenching the coupling of $\Phi_{1,2}$ is equivalent to quenching the masses of $\Phi_\pm$. We can exploit this relation to access regimes that would a priori seem rather artificial.

For instance, consider a global quench where we suddenly change the interaction $g\mapsto g'$ so that $m_-'^2=m^2-g' < 0$. The resulting Hamiltonian density
\begin{equation}
\mathcal{H}_- = \frac{1}{2}\left[\left(\pd_t \Phi_-\right)^2 + \left(\nabla\Phi_-\right)^2 -  |m_-'|^2\Phi_-^2\right],
\label{tachyonicHam}
\end{equation}
has a negative mass term, i.e., it describes a tachyonic dispersion relation, giving rise to a Hamiltonian that is not bounded from below. Operators with these characteristics are usually considered pathological because the associated systems would be intrinsically unstable. However, as we will argue in more detail in later sections, this quenching procedure can still yield a well-defined unitary evolution in the Hilbert space of the original theory. We can then rigorously study the resulting evolution after the quench, even though the driving Hamiltonian lacks a proper ground state. We will call this quenching protocol an unstable tachyonic quench, or more simply, a tachyonic quench.

The most popular way tachyonic field theories are handled in the literature is by adding an extra quartic term $\lambda \Phi^4$. This bounds the Hamiltonian from below while providing a false unstable vaccuum around the quadratic maximum \cite{Peskin}. The most prominent use of this potential is in the symmetry-breaking mechanism that gives mass to gauge fields while preserving gauge invariance \cite{Higgs}. Free tachyonic systems can also be studied by themselves in a rigorous manner \cite{MetaRel, Feinberg, AronsSudarshan, DharSudarshan}. One feature of these latter theories is the restriction of momenta, so that only $|\mathbf{k}|\geq m$ is allowed. We will not need these constraints in the context of this paper because Hamiltonian \eqref{tachyonicHam} will only be used to define an unitary operator dictating the non-equilibrium evolution of a well-defined physical system.

In order to simplify the discussion, we will focus our analysis only on  $\Phi_-$, i.e., on the field whose mass term changes sign after the quench. It should be understood that this is done in the context of the coupled bosonic theory \eqref{coupledBosons} which provides a sensible physical realization.


\section{Quenching the frequency of a simple harmonic oscillator}

The mass quench of a free bosonic QFT has been studied extensively \cite{Tegmark, Sotiriadis1, Sotiriadis2}. One of its main features is that each mode will evolve independently after the quench, with a new frequency determined by the new mass. All the relevant quantities, obtained from the fundamental two-point correlators, can then be reduced to the quenching of independent simple harmonic oscillators (HOs).

Before we study the many-body problem, we must compute the relevant quantities for a single HO. In this context, we want to study the evolution of the ground state of the initial Hamiltonian
\begin{equation}
H_0 =  \frac{\hat p^2}{2} +\frac{1}{2} \omega_0^2 \hat x ^2,
\end{equation}
described by the Gaussian state
\begin{equation}
\braket{x|0} = \left(\frac{\omega_0}{\pi}\right)^{1/4}\exp\left(-\omega_0\frac{x^2}{2}\right),
\end{equation}
under the action of the unitary operator obtained from an inverted quadratic potential
\begin{equation}
U_t = \exp\left[-it\left( \frac{\hat p^2}{2} -\frac{1}{2} \xi^2 \hat x ^2\right)\right] \equiv \exp\left(-it H_1\right).
\label{UnitaryH1}
\end{equation}
We can summarize the protocol as
\begin{equation}
H(t) = \left\{
  \begin{array}{l l}
    H_0, & \,\, t<0,\\
   H_1, & \,\, t\geq 0.
  \end{array} \right. 
\end{equation}
This is the single-body version of the extreme tachyonic quench we want to study in the many-body context.

Operator $H_1$ \eqref{UnitaryH1} is unbounded from below, so it is an ill-defined Hamiltonian. However, we can make sense of it as an operator acting on the original Hilbert space. Consider the ladder operators that diagonalize $H_0$
\begin{equation*}
b = \sqrt{\frac{\omega_0}{2}}\left(\hat x + \frac{i}{\omega_0}\hat p\right).
\end{equation*}
In these variables, we have
\begin{equation*}
H_1 = \frac{\omega_0^2 - \xi^2}{2\omega_0}\left(b^\dg b +\frac{1}{2}\right) - \frac{\omega_0^2 + \xi^2}{4\omega_0}\left((b^\dg)^2 + b^2\right).
\end{equation*}
We see then that $H_1$ is a self-adjoint operator that has a simple and well-defined action on the states of the original Hilbert space. Given that the associated unitary \eqref{UnitaryH1} will have a bounded spectrum, its action is also well-defined for all $t$ \cite{noteH1}.

The propagator for this evolution is given by
\begin{align}
K(x_f,x_i;t) &= \braket{x_f\left|e^{-itH_1}\right|x_i}\nonumber\\
& =\sqrt{\frac{\xi}{2\pi i \sinh(\xi t)}}
\label{IHOpropagator}\\
&\times\exp\left[i\frac{\xi}{2}\frac{\cosh(\xi t)(x_i^2+x_f^2)-2x_i x_f}{\sinh(\xi t)}\right].\nonumber
\end{align}
It corresponds to the familiar propagator of the simple HO after the analytic continuation $\omega \mapsto i\omega=\xi$ \cite{Sakurai}. This remarkable yet natural result can be derived rigorously via the same computational techniques used for the HO \cite{Murayama}.

We can compute the equal-time correlators using propagator \eqref{IHOpropagator}
\begin{align}
\braket{\hat x^2}(t) &= \frac{\omega_0^2 + \xi^2}{4\omega_0 \xi^2}\cosh(2\xi t) - \frac{\omega_0^2 - \xi^2}{4\omega_0 \xi^2},\nonumber\\
\braket{\hat p^2}(t) &= \frac{\omega_0^2 + \xi^2}{4\omega_0 }\cosh(2\xi t) + \frac{\omega_0^2 - \xi^2}{4\omega_0},\\
\braket{\hat x\hat p}(t) &= \frac{\omega_0^2 + \xi^2}{4\omega_0 \xi}\sinh(2\xi t) + \frac{i}{2}.\nonumber
\end{align}
Once again, note that these correspond to analytic continuations of the results obtained from the simple harmonic oscillator \cite{Sotiriadis1, Sotiriadis2}.

One of the most prominent features of these correlators are that they grow exponentially fast. This means that we cannot associate a long-time stationary behavior to the dynamics. In other words, the system will not equilibrate after the quench \cite{GogolinEisert}. If we use the energy levels of the original Hamiltonian as a reference, the expected occupation will evolve as
\begin{align*}
\braket{\hat N}(t) &= \braket{b^\dg b}\\
&=\frac{(\omega_0^2 + \xi^2)^2}{4\omega_0^2 \xi^2}\frac{\cosh(2\xi t)-1}{2}.
\end{align*}
Note that quenching to a free particle ($\xi\to 0$), we obtain a milder growth $\braket{\hat N}\to \omega_0^2 t^2/2$. This implies that the unstable quench can be characterized both by the exponential divergence of the correlators and the occupation of the original energy levels.


\section{UV regularized QFT}

In order to study the many-body version of the extreme tachyonic quench, we will use a standard UV regularized version of the usual free boson QFT. The UV cutoff will be particularly important for the analysis of the scaling of the entanglement entropy after the quench. We will work with a one-dimensional configuration, but all results can be extended to higher dimensions in a straight-forward way (see Appendix A for details).

Consider a set of $N$ harmonic oscillators described by a set of canonical variables $\{\hat p_r, \hat q_r\}$ such that
\begin{equation}
[\hat q_r,\hat p_s] = i\delta_{rs}.
\end{equation}
We define the Hamiltonian
\begin{equation}
H = \frac{1}{2}\sum_{r/a=1}^N\left[\hat p_r^2 + m^2\hat q_r^2 + \Omega^2(\hat q_{r+a} - \hat q_r)^2\right],
\label{UVham}
\end{equation}
where $a=L/N$ is the UV regulatization and we assume periodic boundary conditions. Using the Fourier transform
\begin{equation}
\hat q_r = \frac{1}{\sqrt{N}}\sum_k e^{ikr} \hat q_k,
\end{equation}
(similarly for $\hat p_r$), where momenta $k=\frac{2\pi}{L}n$ are given by $n=0,\pm 1,\cdots,\pm \frac{N-1}{2}$, we have
\begin{align}
H = \frac{1}{2}\sum_k \left[\hat p_k \hat p_{-k} +\omega^2_k \hat q_{k}\hat q_{-k}\right]
\label{HamMomentum}
\end{align}
where
\begin{equation}
\omega^2_k = 4\Omega^2\sin^2\left(\frac{ka}{2}\right) + m^2.
\end{equation}
Note that if we want a well-defined continuum limit
\begin{equation}
aN=L \text{  fixed}, \quad N\to \infty, a\to 0,
\label{thermoLimit}
\end{equation}
with a relativistic dispersion relation
\begin{equation}
\omega_k^2 \approx a^2\Omega^2 k^2 + m^2,
\end{equation}
the associated speed of light will be $c=a\Omega$.

Hamiltonian \eqref{HamMomentum} implies that we can associate an independent HO to each mode $k$. If we quench the mass of the system, it will be equivalent to changing the frequencies from
\begin{equation}
\omega_{0k} = \sqrt{4\Omega^2\sin^2\left(\frac{ka}{2}\right)+m_0^2}
\end{equation}
to
\begin{equation}
\omega_{k} = \sqrt{4\Omega^2\sin^2\left(\frac{ka}{2}\right) - m^2} \equiv i\xi_k.
\label{omega_k}
\end{equation}
We see that the oscillators will have two possible dynamics:

a) \textit{Stable modes:} If $m\leq 2\Omega|\sin(ka/2)|$, the final frequency $\omega_k$ will be real. In that case the evolution will be like a simple quench from a harmonic Hamiltonian to another. These modes will behave qualitatively as regular mass quenches \cite{Sotiriadis1, Sotiriadis2}. In particular, we expect this modes to equilibrate. Sotiriadis, Calabrese and Cardy \cite{Sotiriadis1} obtained an effective temperature for the equilibration of these modes by comparing it to the Matsubara propagator. It can be written as
\begin{equation}
\beta_{\text{eff}} = \frac{1}{\omega_k}\log\left(\frac{\left(\omega_k+\omega_{0k}\right)^2}{\left(\omega_{0k}-\omega_k\right)^2}\right).
\end{equation}
If we assume that $\omega_{0k}\gg \omega_k$, we can find a simplified relation
\begin{equation*}
\beta_{\text{eff}} \approx \frac{4}{\omega_{0k}}\left(1+\frac{\omega_k^2}{3\omega_{0k}^2}\right).
\end{equation*}
The momentum dependence is consistent with the fact that integrable systems (such as free bosons) do not equilibrate globally after a quench. This is due to the presence of a macroscopic number of conserved quantities. Thermal ensembles should then be replaced by generalized Gibbs ensembles (GGE) that take into account all the independent integrals of motion \cite{GogolinEisert, GGE}.


b) \textit{Unstable modes:} If $m>2\Omega|\sin(ka/2)|$, we expect these modes to behave according to the unstable quench we did for the simple HO in the previous section. It follows that they not only do not equilibrate, but also explode exponentially fast. As we will see in later sections, these modes (the low-frequency ones) will dominate the long-time behavior of the system. This makes it impossible to define an stationary ensemble that captures the qualitative physical properties after the quench.


\section{A note on quenching regimes}

For simplicity, we will be working in the so-called deep quench limit \cite{Sotiriadis2}. It is characterized by an extremely massive initial state $m_0\gg m$, so that the initial correlation lenght vanishes and we have an effective product state. In this limit, all the correlations formed after the quench are due to the evolution, allowing for a clear analysis. 

In our numerical calculations, we will set $a=1$ and $\Omega=\sqrt{N}$. We will use the associated speed of light $c=a\Omega$ to give the correct relativistic scale to the time direction. Note that if $m\gg \Omega$, the harmonic oscillators will basically decouple and the evolution will be similar to $N$ independent HOs. Also, the corresponding continuum limit will be an infinitely massive theory, which is very limiting. We will then prefer $m\ll\Omega$, so that the dynamical features are also prevalent. 



\section{Correlators after the quench}

Being a Gaussian state evolving according to a Gaussian unitary, we expect the state to remain Gaussian for all times after the quench. That implies that everything will be determined by the fundamental two-point functions. These correlators can be written as
\begin{align}
\braket{\hat q_r \hat q_s}(t) &= \frac{1}{N}\sum_k e^{ik(r-s)} C^{(qq)}_k(t),\nonumber\\
\braket{\hat p_r \hat p_s}(t) &= \frac{1}{N}\sum_k e^{ik(r-s)} C^{(pp)}_k(t),\\
\braket{\hat q_r \hat p_s}(t) &= \frac{1}{N}\sum_k e^{ik(r-s)} C^{(qp)}_k(t),\nonumber
\end{align}
where
\begin{align}
C^{(qq)}_k(t) &=\frac{\omega_{0k}^2 + \omega_k^2}{4\omega_{0k} \omega_k^2} -\frac{\omega_{0k}^2 - \omega_k^2}{4\omega_{0k} \omega_k^2}\cos(2\omega_k t),\nonumber\\
C^{(pp)}_k(t) &=  \frac{\omega_{0k}^2 + \omega_k^2}{4\omega_{0k}} + \frac{\omega_{0k}^2 - \omega_k^2}{4\omega_{0k} }\cos(2\omega_k t),\\
C^{(qp)}_k(t) &=\frac{\omega_{0k}^2 - \omega_k^2}{4\omega_{0k} \omega_k}\sin(2\omega_k t) + \frac{i}{2}.\nonumber
\end{align}
Note that these functions are even in $\omega_k$, so there is no branch cut in the $k$ complex plane due to the square root in definition \eqref{omega_k}. Also, in the case $\omega_k^2$ is negative, both possible imaginary roots will give the same functions, so there is no ambiguity. 

In the deep quench limit, there are some simplifications. The initial mass scale $m_0$ decouples in the correlators and we have
\begin{align}
\braket{\hat q_r \hat q_s}(t) &= \frac{m_0}{4N}\sum_k e^{ik(r-s)}\left[\frac{1-\cos(2\omega_k t)}{\omega_k^2}\right] ,\nonumber\\
\braket{\hat p_r \hat p_s}(t) &= \frac{m_o}{4N}\sum_k e^{ik(r-s)}\left[1+\cos(2\omega_k t) \right] ,\label{quenchCorr}\\
\braket{\hat q_r \hat p_s}(t) &=  \frac{m_0}{4N}\sum_k e^{ik(r-s)}\left[\frac{\sin(2\omega_k t)}{\omega_k}\right] + \frac{i}{2}\delta_{rs}.\nonumber
\end{align}
Note that these correlators are valid for $|r-s|, t\gg 1/m_0$.


\section{Causal structure after the quench}

Even though tachyonic theories are usually associated to superluminal particles, they are better undestood as instabilities in causal theories \cite{AKS, Sen, Bruneton}. Propagation is still governed by causal Green functions, in such a way that information cannot move faster than the speed of light. In the context of the tachyonic quench, we can explicitly show that the causal structure is preserved during the evolution.

\begin{figure}[h]
  \centering

\includegraphics[width=0.9\linewidth]{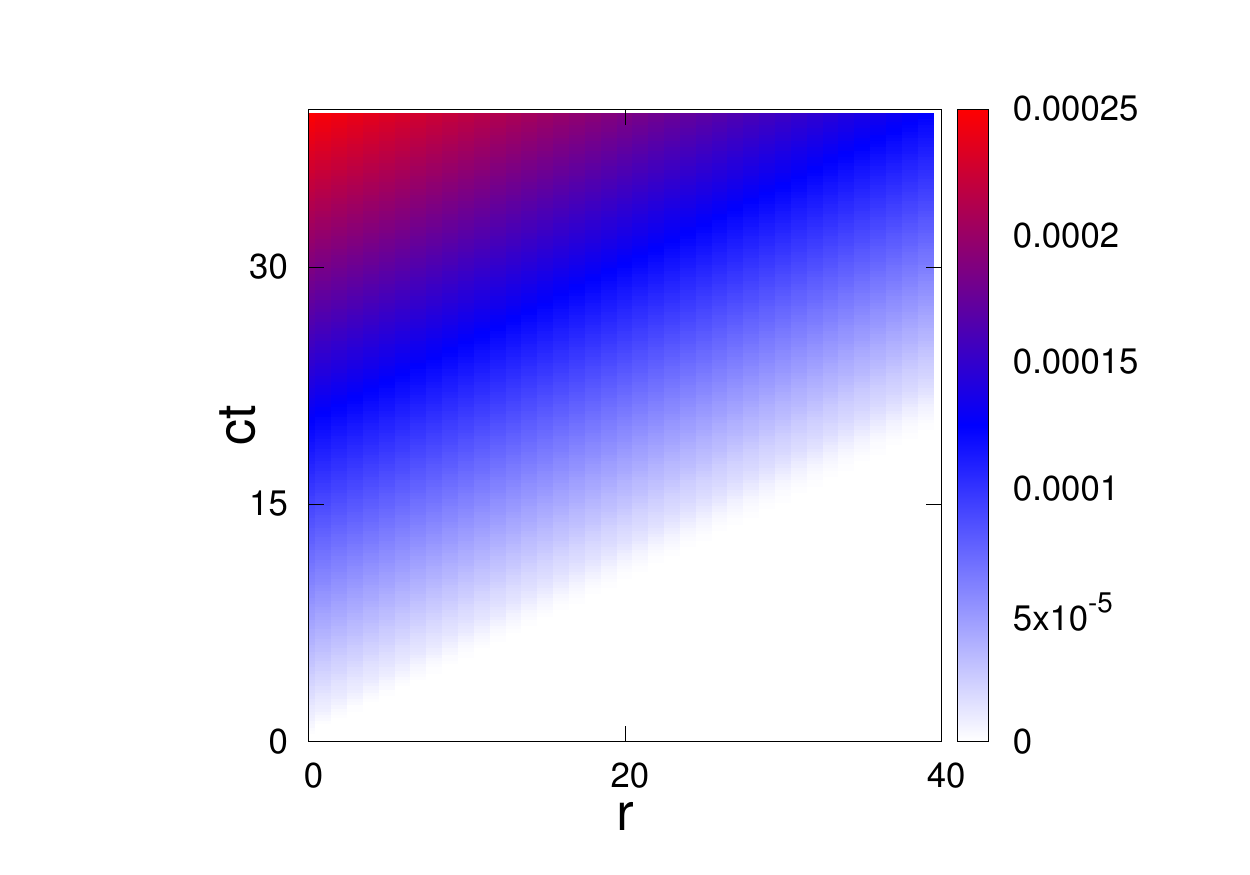}

    \caption{Density plot of the correlator $\braket{\hat q_r \hat q_0}(t)/m_0$ in the deep quench limit for $m=1$, and $N=40000$. Note that the causal structure is strictly preserved.}
    \label{correlator}
\end{figure}

First, consider the UV regulated Hamiltonian \eqref{UVham}. Being a local lattice system, the dynamics can be studied using Lieb-Robinson bounds \cite{LRoriginal}. For quadratic bosonic systems, the maximum speed of propagation can be obtained from the couplings of the Hamiltonian \cite{Nachtergaele, LRbosons} . Following \cite{LRbosons}, we obtain (see Appendix B for details)
\begin{equation}
v_\text{LR} = ea\max |\omega_k| = ea\sqrt{\left|4\Omega^2-m^2\right|}\approx 2ec.
\end{equation}
Commutators of operators that are "space-like" separated with respect to $v_\text{LR}$ will be strongly suppressed. In particular, if $2r>v_\text{LR}t$, we have 
\begin{align}
\norm{[\hat q_r(t),\hat q_0]} \leq \frac{1}{\max |\omega_k|}  \frac{e^{-2(r/a)\log(2r/v_\text{LR}t)}}{\sqrt{r/a}(1-\frac{v_\text{LR}t}{2r})},\\
\norm{[\hat p_r(t),\hat p_0]} \leq \max |\omega_k| \frac{e^{-2(r/a)\log(2r/v_\text{LR}t)}}{\sqrt{r/a}(1-\frac{v_\text{LR}t}{2r})},\nonumber
\end{align}
where $\hat A_r(t) = U_t^\dg \hat A_r U_t$ and $\norm{\cdots}$ is the operator norm. As we see in Fig. \eqref{correlator}, correlators agree with this causal light-cone. This is consistent with the associated speed of light (the extra $e$ is due to the fact that this bound is obtained for general lattices \cite{LRbosons}).

The structure of the light-cone in the correlators is more explicit in the continuum limit, where causality is also preserved. If we set $\Omega = \frac{1}{a}$, we have
\begin{align}
\omega_{0k}^2 \to k^2 + m_0^2,\quad \omega_{k}^2 \to k^2 - m^2,
\end{align}
as $a\to 0$ and
\begin{equation}
\frac{\braket{\hat q_r \hat q_s}(t)}{a} \to \frac{m_0}{4} \int \frac{dk}{2\pi} e^{ik(r-s)}\left[\frac{1-\cos(2\omega_k t)}{k^2 - m^2}\right].
\end{equation}
Note that this propagator will vanish for $r>2t$. This follows from evaluating the integral by taking the principal value around $k=\pm m$ and closing the integration contour in the upper half of the complex $k$ plane. (A similar argument is used in \cite{AKS} to relate the dynamics of a tachyonic field to instabilities in the theory. See also \cite{Sotiriadis2} for an analysis of the case $m\to 0$.)


\section{Entanglement and mutual information growth after the quench}

Entanglement entropy (EE) is perhaps one of the simplest measures that characterize the information interdependence between subsystems. Given that after a quantum quench a system can survey a big portion of the Hilbert space, the evolution of the EE of a fixed subsystem can be used as proxy to understand in an unified way the formation of new correlations.

\begin{figure}[h]
\centering
\subfigure[]{
   \includegraphics[width=0.85\linewidth]{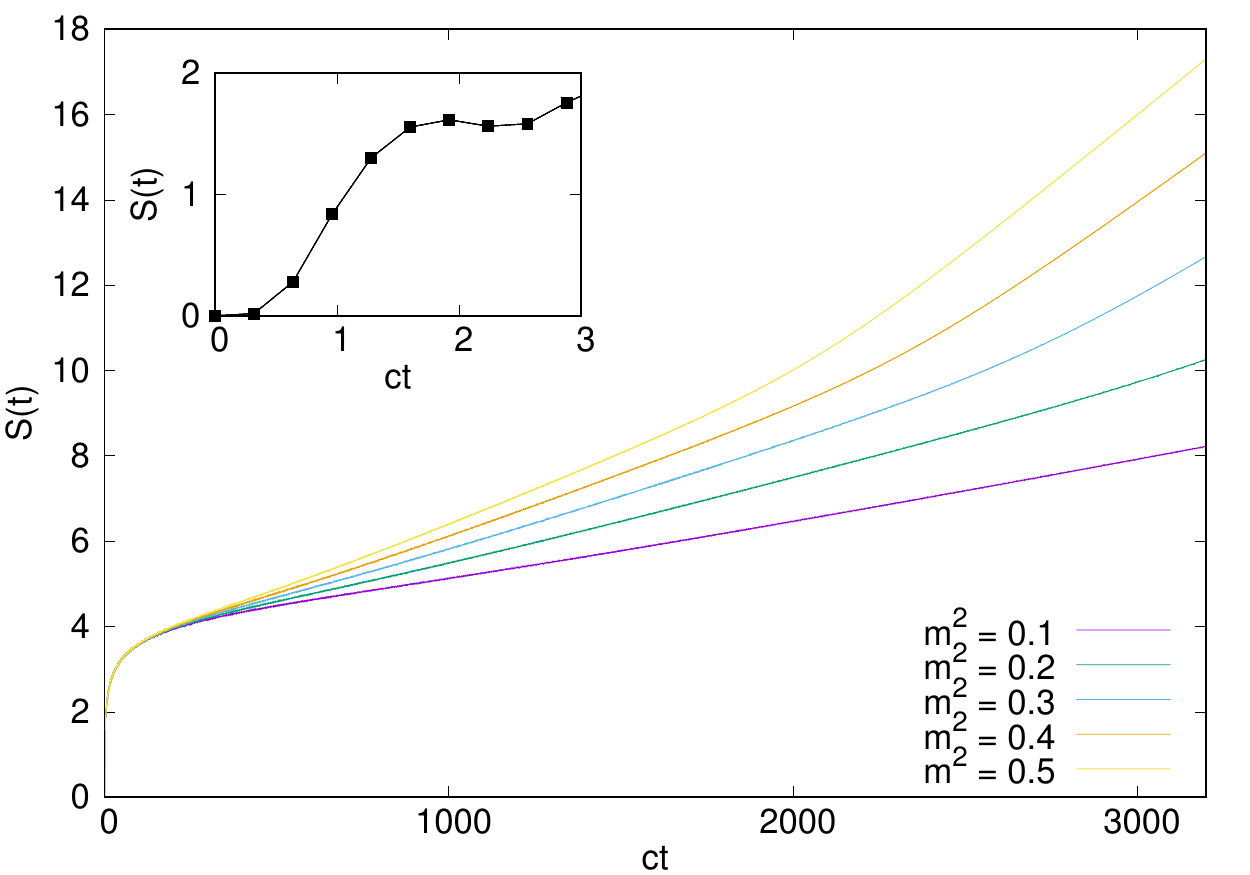}
   }
\subfigure[]{
   \includegraphics[width=0.85\linewidth]{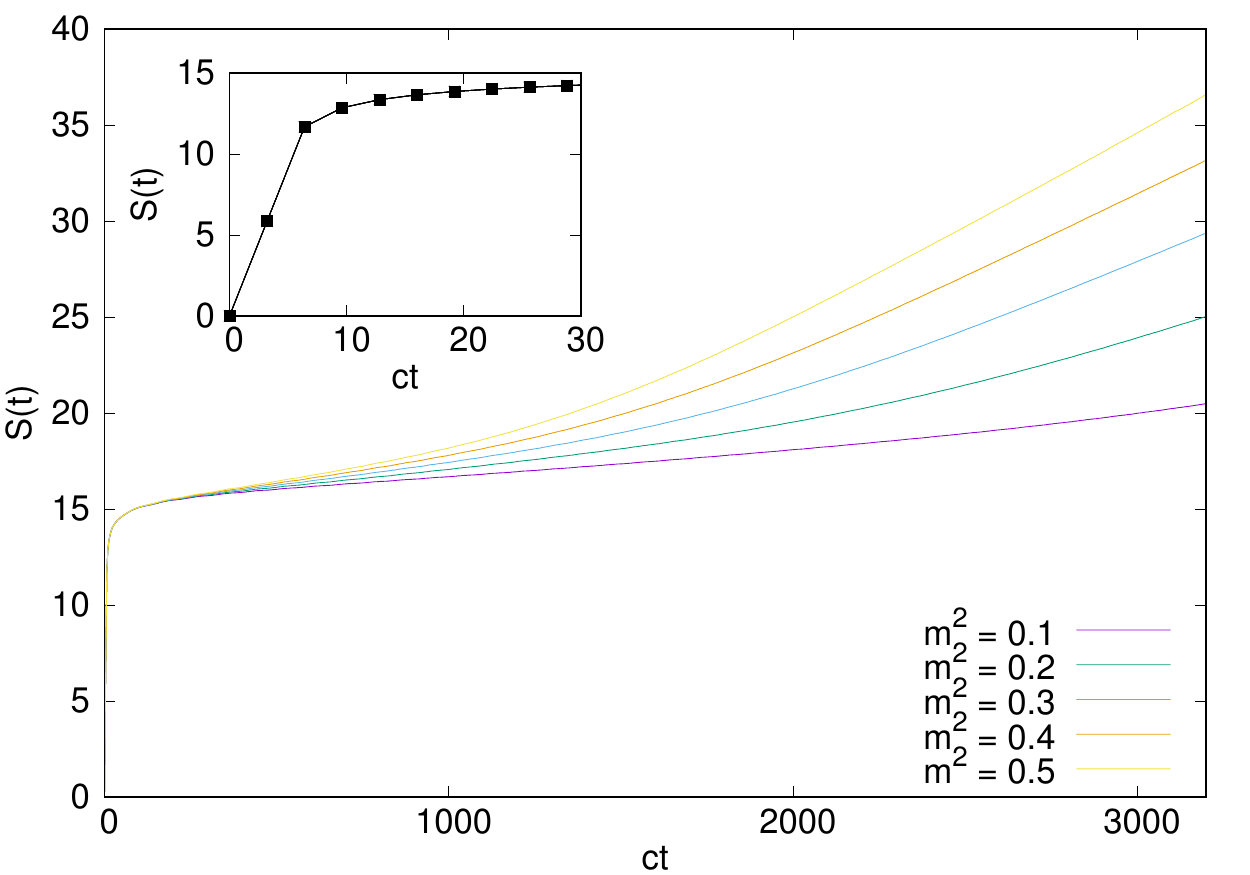}
}
    \caption{Evolution of the EE of a block of size (a) $L=1$ and (b) $L=10$ after a tachyonic quench with $m_0=1000$, $N=40000$, and $m^2=0.1,\cdots,0.5$. Observe that the EE starts growing in a linear fashion for $mt\gg L$. This implies that for small $m$ the growth will be logarithmically slow until the right scale is reached. \emph{Inset:} For short times, the growth is approximately linear and independent of $m$.}
    \label{EEL}
\end{figure}

Observables in Gaussian systems are completely determined by the covariance matrix
\begin{equation}
\Gamma_{nm} = \text{Re}\braket{\hat r_n \hat r_m},
\end{equation}
where $\mathbf{\hat r} = (\hat q_1,\cdots,\hat q_N,\hat p_1,\cdots,\hat p_N)$. In particular, all information about a subsystem $A$, composed of sites $\{i_1,\cdots,i_L\}$, can obtained from the $2L\times 2L$ submatrix
\begin{equation}
\Gamma^{A}_{nm} = \Gamma_{i_n i_m},\quad \Gamma^{A}_{n+L,m} = \Gamma_{i_n + N, i_m}, \text{ etc.}.
\end{equation}
The EE of subsystem $A$ can be computed from the associated symplectic eigenvalues $\{\sigma_n|n=1,\cdots,L\}$, where $\sigma_n\geq 1/2$. These correspond to the positive spectrum of $i\Gamma^A\Omega_\text{sym}$, where $\Omega_\text{sym}$ is the symplectic matrix
\begin{equation}
\Omega_\text{sym} = \begin{pmatrix}0 & \id_{L} \\ -\id_L & 0\end{pmatrix}.
\end{equation}
The EE is given by the formula \cite{EntropyBosons, EntropyBosonsReview, Demarie}
\begin{equation}
S_A = \sum_{n=1}^L \left(f\left(\sigma_n+\frac{1}{2}\right) -f\left(\sigma_n-\frac{1}{2}\right) \right),
\end{equation}
where $f(x)=x\log(x)$. Note that for very large symplectic eigenvalues $\sigma_n\gg 1$, we have $f(\sigma_n+\frac{1}{2}) -f(\sigma_n-\frac{1}{2})\approx \log(\sigma_n)+1$.

\begin{figure}[h]
  \centering
\subfigure[]{
    \includegraphics[width=0.9\linewidth]{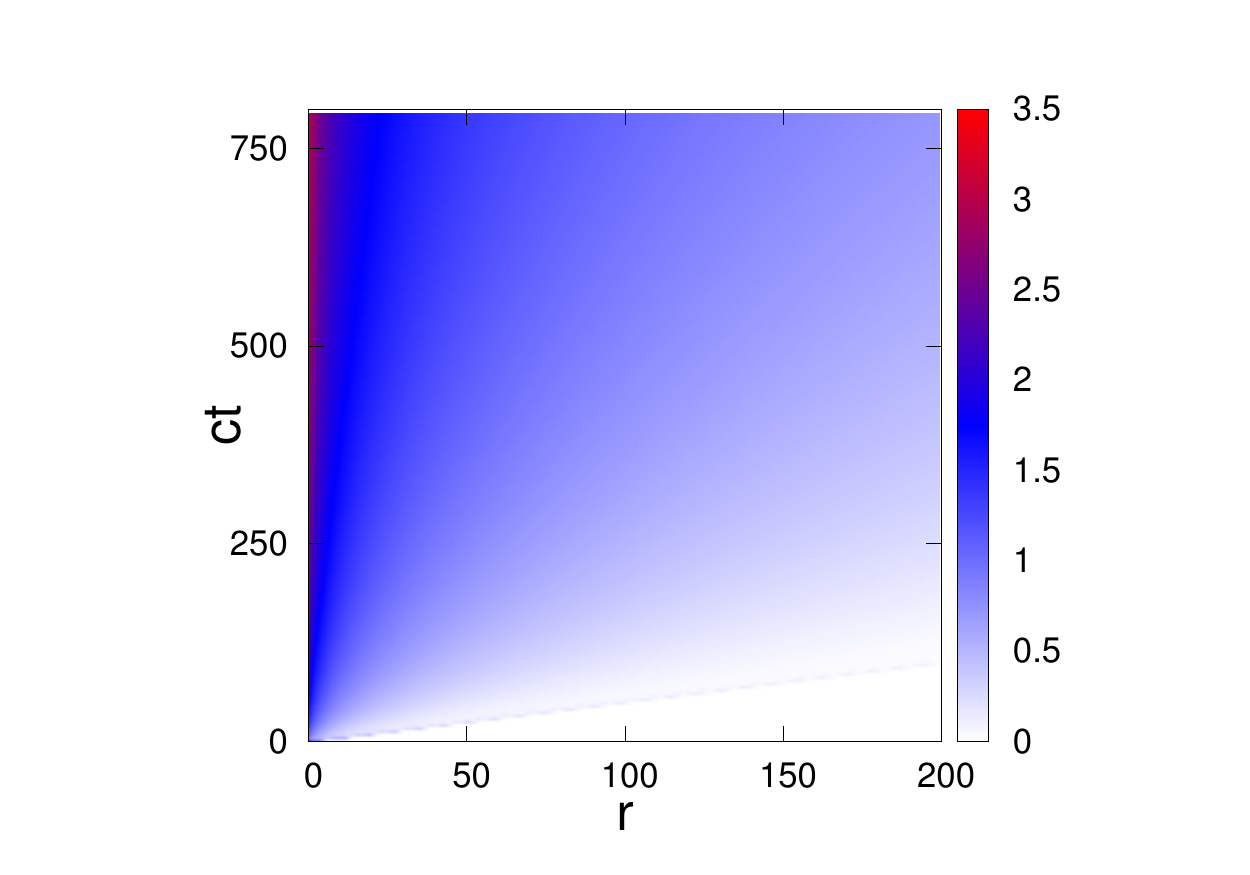}
}
\subfigure[]{
  \includegraphics[width=0.9\linewidth]{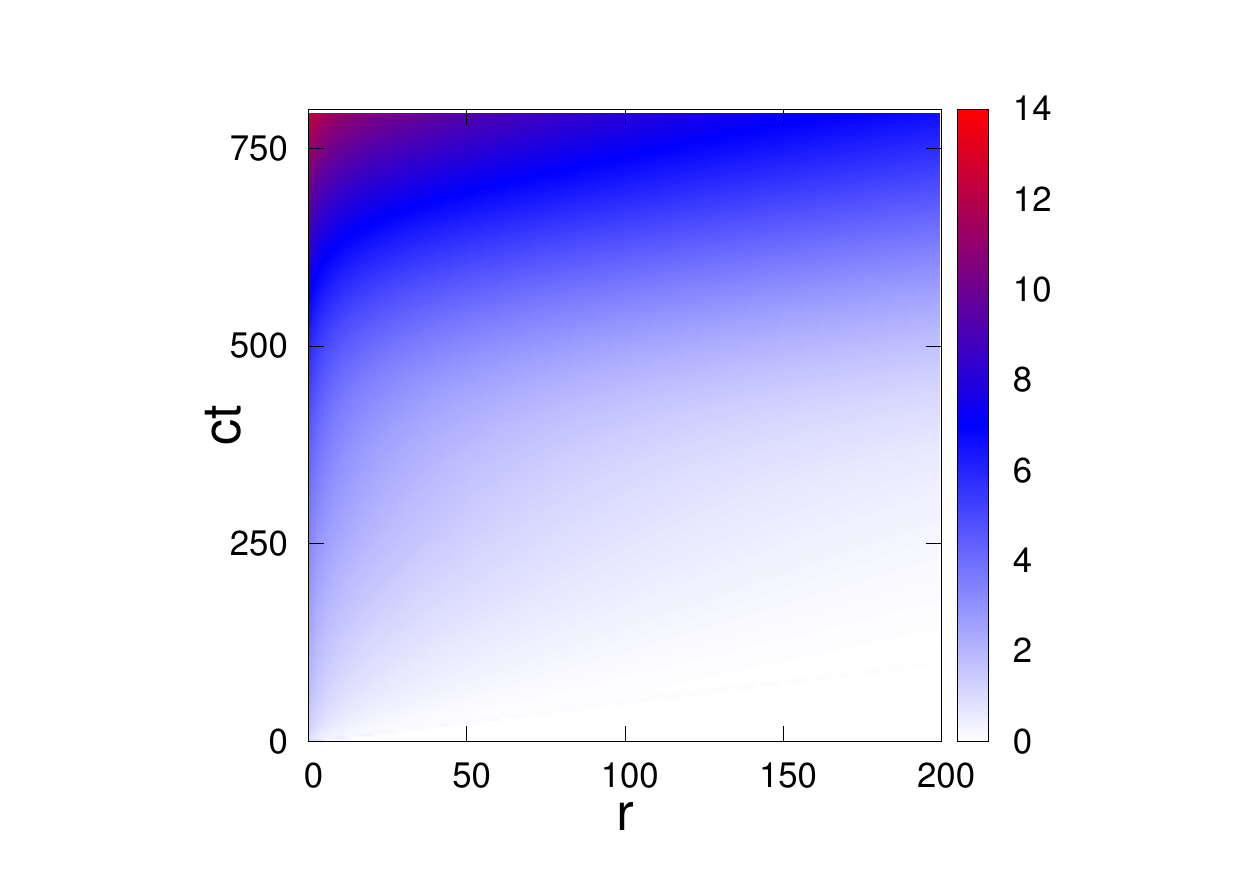}
\label{fid-x0}
}
    \caption{Contour plot for the evolution of the mutual information of two blocks consisting of 3 contigous sites each separated by a distance $r$ for $m_0=1000$ and $N=40000$ after (a) a massless quench $m=0$ and (b) a tachyonic quench with $m=2$.}
    \label{mInforContour}
\end{figure}

Now, before we discuss the evolution of the EE after an unstable tachyonic quench, let us consider the long-time evolution of the correlators. If we take the continuum limit in eq. \eqref{quenchCorr}, we have
\begin{equation}
\frac{\braket{\hat q_r \hat q_s}(t)}{a} \to \int\frac{dk}{2\pi}e^{ik(r-s)}C_k^{(qq)}(t).
\label{intQQ}
\end{equation} 
For large times $t\gg 1/m$, we have
\begin{equation}
C_k^{(qq)}(t) \approx \frac{m_0}{4\xi_k^2}\exp(2\xi_k t),
\end{equation}
where we also used the deep quench limit. The integrand in \eqref{intQQ} will be sharply peaked around $k=0$, so we can do a steepest descent approximation and obtain
\begin{equation}
\log\left(\frac{\braket{\hat q_r \hat q_s}(t)}{a}\right) \to 2mt + \mathcal{O}(\log(mt)).
\label{leadingQQ}
\end{equation}
Similarly for $\braket{\hat p_r \hat p_s}$ and $\text{Re}\braket{\hat q_r \hat p_s}$. This implies that all the elements in $\Gamma^A$ have the same exponentially divergent factor. We expect then that for $ct\gg L$, we have
\begin{equation}
S_A \sim 2mLt + \mathcal{O}(\log(mt )),
\label{linEntr}
\end{equation}
where the volumetric factor $L$ comes from the number of symplectic eigenvalues. This behavior is universal. Note that this result is also valid for higher dimensions because the leading divergence of the corresponding correlators will also be an exponential \eqref{leadingQQ}.

In Fig. \eqref{EEL} we see the evolution of $S^A(t)$ after an unstable tachyonic quench for blocks of sizes $L=1$ and $L=10$.  We see there is a short transient time $t_\text{trans}\sim L/\Omega$ during which information propagates ballistically independent of $m$ (see inset). This is the same type of propagation exhibited in the quench dynamics of conformal field theory \cite{LCpropagation, CardyCalabrese}. However, the instability of the driving Hamiltonians takes over and we end up with a linear growth \eqref{linEntr}.

\begin{figure}[h]
  \centering

\includegraphics[width=0.9\linewidth]{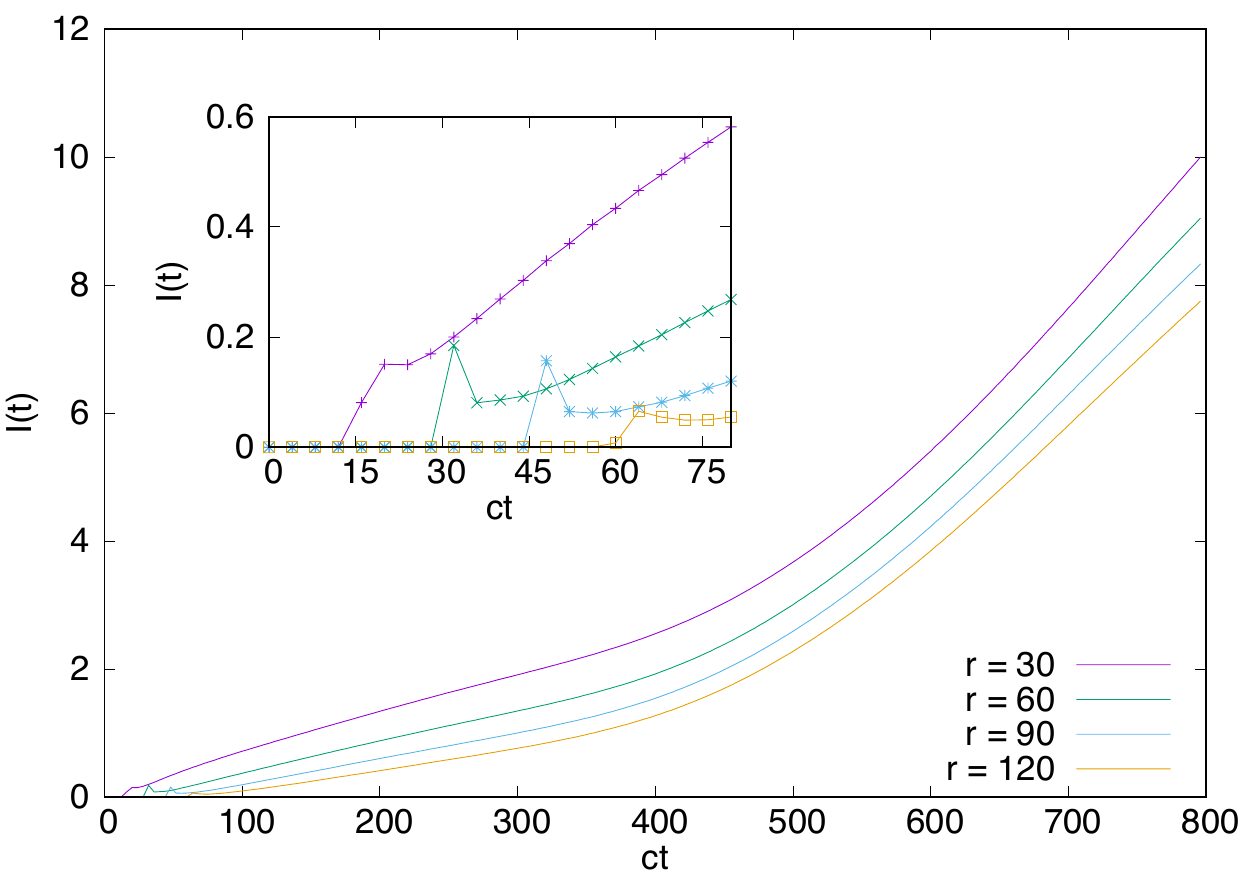}
    \caption{Evolution of the mutual information of two blocks consisting of 3 contigous sites each separated by distances $r=30,60,90,120$ after a tachyonic quench with $m_0=1000$, $m=2$ and $N=40000$.}
    \label{EvolMInfo}
\end{figure}

We can gain further insight about the growth of correlations by studying the mutual information between subsystems. For two disjoint regions $A$ and $B$, it is defined as
\begin{equation}
I_{AB} = S_A + S_B - S_{A\cup B}.
\end{equation}
This can be used as a measure of the amount of correlation between $A$ and $B$, even serving as an upper bound for normalized two-point functions \cite{MutualInfo}.

Figure \eqref{mInforContour} illustrates the evolution of the mutual information after a massless quench ($m^2=0$) and a tachyonic quench for two blocks $A=\{1,2,3\}$ and $B=\{r+1,r+2,r+3\}$. Both display an unbounded growth with a sharp light-cone. However, the massless quench only grows logarithmically in time, while the tachyonic grows linearly. Figure \eqref{EvolMInfo} shows this evolution for fixed distances. Note how the mutual information starts growing after $t=r/2c$, as imposed by the causal structure. In the same fashion as the EE, this implies that the massless modes dominate the dynamics for short times, but the instability ends up leading the behavior.

The linear growth of the mutual information after a tachyonic quench is remarkable because it shows that the divergence in the EE is not a simple artifact due to the size of the local Hilbert space. Also, note that the growth cannot follow from the asymptotic behavior \eqref{linEntr} because that volumetric contribution cancels.


\section{Possible connections to physical systems}

Given the instabilities of the setting we have described, it is easy to see that any physical realization will have severe constraints. However, it is possible to find regimes that can approximate this type of quenches for certain periods of time. In particular, if we include higher order terms in the Hamiltonian to constrain the instabilities, we expect that tachyonic behavior can be seen for timescales that are shorter than the ones that bound the dynamics. 

Consider first an $O(3)$ non-linear $\sigma$-model (NLSM) described by the Hamiltonian \cite{Affleck}
\begin{equation}
H = \int dx \left[g^2\left(\vec{\ell}-\frac{\theta}{4\pi}\pd_x\vec{\phi}\right)^2+\frac{1}{g^2}\pd_x\vec{\phi}^2\right],
\end{equation}
where $\vec{\phi}=(\phi_1,\phi_2,\phi_3)$ is contrained by $\vec{\phi}^2=1$ and $\vec{\ell} = \vec{\phi}\times \pd_t\vec{\phi}$. If we expand around a classically ordered state $\phi_3^2\approx 1$, we can approximate the system as two independent bosonic fields described by the Lagrangian density
\begin{equation}
\mathcal{L} = \frac{1}{g}\left[(\pd_\mu\phi_1)^2 + (\pd_\mu\phi_2)^2\right].
\end{equation}
As we saw in Sect. II, we can obtain a tachyonic sector by adding a coupling of the form $\phi_1\phi_2$. We can justify this microscopically by using the well-known map that relates the $O(3)$ NLSM to the antiferromagnetic Heisenberg spin chain \cite{Affleck}
\begin{equation}
S^a_{2n}= s \phi_a(x) + \ell_a(x), \quad S^a_{2n+1} = - s \phi_a(x) + \ell_a(x),
\end{equation}
where $S^a_n$ are the spin operators ($a=1,2,3$), $s\gg 1$ is the total (local) spin, and $x=2n+\frac{1}{2}$. Setting $\phi_\pm = (\phi_1\pm \phi_2)/\sqrt{2}$, we have
\begin{align}
m^2\sum_x (\phi_+^2 - \phi_-^2)& (x) \nonumber\\
= m^2\sum_n &\left[\{S^x_{2n},S^y_{2n}\} + \{S^x_{2n+1},S^y_{2n+1}\} \right.\\
&\left. - 2S^x_{2n}S^y_{2n+1} - 2S^y_{2n}S^x_{2n+1}\right].\nonumber
\end{align}
Note that this interaction term will give rise to an unstable Hamiltonian in the limit $s\to \infty$. Yet, the normalization constraint $\vec{\phi}^2=1$ will eventually bound the dynamics as soon as the approximation $\phi_3^2\approx 1$ breaks down.

Consider now a free boson with a quartic interaction
\begin{equation}
\mathcal{H} = \frac{1}{2}\left[\left(\pd_t \phi\right)^2 + \left(\nabla\phi\right)^2 +  m^2\phi^2 + \frac{1}{4!}\lambda\phi^4\right].
\end{equation}
By considering a self-consistent substitution using the Hartree-Fock approximation \cite{phi4}
\begin{equation}
\phi^4 \to -3\braket{\phi^2}^2 + 6\braket{\phi^2}\phi^2,
\end{equation}
we can define an effective mass \cite{Sotiriadis2}
\begin{equation}
m_\text{eff}^2 = m^2 + \frac{\lambda}{2}\sum_k \braket{\phi_k^2}.
\end{equation}
If the mass is tachyonic $m^2\mapsto -m^2$, heuristically we expect the interaction to stop the instability after a characteric time $t_\text{stable}$ given by
\begin{equation}
\sum_k \braket{\phi_k^2}(t_\text{stable}) \sim 2\frac{m^2}{\lambda}.
\end{equation}
As we discussed in previous sections, the correlators after a tachyonic quench develop an exponential growth. This implies that the interaction term $\lambda \phi^4$ must be very small compared to the rest of the characteristic terms in order to approximate a tachyonic evolution.


\section{Discussion}

Tachyonic quenches are an exotic alternative for the study of bosonic field theories out of equilibrium. As we saw in all of the computed physical quantities, the causal structure is made manifest in a fashion similar to critical quenches. For short times, the evolution of observables may even be indistinguishable. However, tachyonic quenches are unstable and exponential divergences end up dominating the behavior. In particular, the low-frequency (unstable) modes $c|k|<m$ are exponentially driven and cannot equilibrate.

Tachyonic (or more generally, unstable) quenches can be used as an intermediate preparation step for many-body states. If the driving Hamiltonian \eqref{UnitaryH1} is only used for a fixed time $T$, the resulting state will be highly excited in the low frequency modes while approximately thermal for the high frequency ones. This sets a sharp cut-off around the tachyonic mass, allowing for a dynamical separation of scales. Remarkably, a similar mechanism can be found in the statistical physics of fluids and interfaces, in processes described by the Kuramoto-Sivashinski formula \cite{turbulence}.

In order to obtain similar dynamics in other physical systems, the driving Hamiltonian must have some sort of instability. However, being a pathological feature that is generally avoided, this unboundedness is usually absent by construction in physical realizations. For example, extending these constructions to fermions is not straight forward, at least using free Hamiltonians. This is due to the symmetries of the energy spectrum, that is traditionally interpreted as a Dirac sea. Spin systems could be used, but they would need very large quantum numbers, so that they can be close to the quantum rotor limit. In this case, the parameters must be chosen so that the characteristic times of the instability are shorter than the one imposed by the lower bound of the spectrum of the driving Hamiltonian.

Realizations may also use a $\lambda \phi^4$ potential. However, in order to exploit this resource at its fullest, the parameters must be chosen so that the quartic term becomes relevant after the exponential divergence becomes manifest.

Further work is needed to understand other types of unstable quenches. Interacting terms that produce unbounded Hamiltonians can for example be fine-tuned to obtain other types of long-time divergences. Characterizing approximate realizations using truncated local Hilbert spaces may also provide an interesting setting for future experiments using optical lattices.


\section*{Acknowledgments}

We would like to thank G. Mussardo and A. Botero for useful discussions. This work is supported by the Spanish Research Agency (Agencia Estatal de Investigaci\'on) through the grant IFT Centro de Excelencia Severo Ochoa SEV-2016-0597, and funded by Grant No. FIS2015-69167-C2-1-P from the Spanish government, and QUITEMAD+ S2013/ICE-2801 from the Madrid regional
government. SM is supported by the FPI-Severo Ochoa Ph.D. fellowship No. SVP-2013-067869.


\section*{Appendix A: Basic results for higher dimensions}

Hamiltonian \eqref{UVham} can be written in arbitrary dimensions
\begin{equation}
H = \frac{1}{2}\sum_{r}\left[\hat p_r^2 + m^2\hat q_r^2 + \sum_{s=1}^d\Omega^2(\hat q_{r+a_s} - \hat q_r)^2\right],
\label{UVHamAnyD}
\end{equation}
where $a_s$ is a displacement in the lattice by one site along the $s$-th dimension. It is diagonalized in the same way
\begin{align}
H = \frac{1}{2}\sum_\mathbf{k}\left[\hat p_\mathbf{k} \hat p_{-\mathbf{k}} +\omega^2_\mathbf{k}\, \hat q_\mathbf{k}\hat q_{-\mathbf{k}}\right],
\end{align}
with the dispersion relation
\begin{equation}
\omega_\textbf{k}^2 = m^2 + 4\Omega^2\sum_{s=1}^d \sin^2\left(\frac{k_s a}{2}\right),
\end{equation}
where $\mathbf{k}=(k_1,\cdots,k_d)$ and
\begin{equation}
k_s = \frac{2\pi}{L_s}n, \qquad n=0,\pm 1,\cdots, \pm \frac{N_s-1}{2}.
\end{equation}
Note that in the continuum limit \eqref{thermoLimit}, the associated speed of light is still $c=a\Omega$.


\section*{Appendix B: Lieb-Robinson bound for harmonic systems}

We will summarize some results about Lieb-Robinson bounds for harmonic systems following \cite{LRbosons}. We will focus only on the type of Hamiltonians discussed in this paper and on commutators of $\{\hat q_i\}$ for the sake of concreteness.

Consider the Hamiltonian
\begin{equation}
H = \frac{1}{2}\sum_{i,j} \left(\hat q_i X_{ij}\hat q_j + \hat p_i P_{ij} \hat p_j\right),
\end{equation}
where $X,P\in \mathbb{R}^{N\times N}$ are symmetric matrices. For simplicity, we will assume that $P_{ij}=\delta_{ij}$ and a nearest-neighbor Hamiltonian
\begin{equation}
X_{ij} = 0, \qquad \text{ for } d(i,j)>1,
\label{localX}
\end{equation}
where $d(i,j)$ is the graph-theoretical distance. (Note that we are not assuming that the lattice has a particular spatial dimension.) If we define $\hat q_n (t) = e^{itH}\hat q_n e^{-itH}$, we have \cite{LRbosons}
\begin{equation}
i[\hat q_n(t), \hat q_m] = \sum_{s=0}^\infty \frac{(-1)^s t^{2s+1}}{(2s+1)!}(X^s)_{nm}\equiv C^{qq}_{nm}(t)\id.
\end{equation}
This identity holds even if $H$ is unbounded from below because it is obtained using commutators of $\hat q_i$ and $\hat p_i$ with $\exp(\alpha H)$ via the Baker-Hausdorff formula \cite{LRbosons}.

Coefficient $C^{qq}_{nm}(t)$ can be bounded using the locality condition \eqref{localX}, so that
\begin{equation}
\left|C^{qq}_{nm}(t)\right| \leq \frac{1}{\sqrt{\norm{X}}} \sum_{s=0}^\infty\frac{\tau^{2s+2d(i,j)+1}}{(2s+2d(i,j)+1)!},
\end{equation}
where $\norm{X}$ is the operator norm and
\begin{equation}
\tau = \sqrt{\norm{X}}t.
\end{equation}
For $e\tau<2c$, for $c\in\mathbb{Z}^+$, we have
\begin{equation}
\sum_{s=c}^\infty \frac{\tau^{2s}}{(2s)!}\leq \frac{(e\tau/2c)^{2c}}{\sqrt{2c}(1-(e\tau/2c)^2}.
\label{boundIdentity}
\end{equation}%

Now, being a finite symmetric matrix, $\norm{X}$ correspond to its largest eigenvalue in absolute value. If we assume Hamiltonian \eqref{UVHamAnyD} with $4\Omega^2>|m^2|$, it is easy to see that
\begin{equation}
\norm{X} = \max_\mathbf{k}|\omega_\textbf{k}| = 4d\Omega^2+m^2.
\label{normX}
\end{equation}
Using \eqref{boundIdentity} and \eqref{normX}, we have that the Lieb-Robinson bound for Hamiltonian \eqref{UVHamAnyD} is
\begin{equation}
v_\text{LR} = ea\sqrt{4d\Omega^2+m^2}.
\end{equation}
Note that this result holds for $m^2\mapsto -m^2$. Also, the bound will always be larger than the speed of light $c=a\Omega$ as long as $\Omega \gg |m|$.



\begin{thebibliography}{99}
\providecommand{\url}[1]{\texttt{#1}}
\providecommand{\urlprefix}{URL }
\providecommand{\eprint}[2][]{\url{#2}}

\bibitem{Experiments1}
M. Greiner et al.,
Nature \textbf{419}, 51 (2002);
S. Hofferberth et al.,
Nature \textbf{449}, 324 (2007);
M. Cheneau, et al.,
Nature \textbf{481}, 484 (2012);
T. Langen, et al.,
Nat. Phys. \textbf{9}, 640 (2013);
P. Jurcevic, et al.,
Nature \textbf{511}, 202 (2014).

\bibitem{CardyCalabrese1}
P. Calabrese and J. Cardy,
J. Stat. Mech. (2007) P06008.

\bibitem{GogolinEisert}
C. Gogolin and J. Eisert,
Rep. Prog. Phys. \textbf{79}, 056001 (2016).

\bibitem{MetaRel}
O. M. P. Bilaniuk, V. K. Deshpande, and E. C. G. Sudarshan,
Am. J. Phys. \textbf{30}, 718 (1962).

\bibitem{Feinberg}
G. Feinberg,
Phys. Rev. \textbf{159}, 1089(1967).

\bibitem{AronsSudarshan}
M. E. Arons and E. C. G. Sudarshan,
Phys. Rev. \textbf{173}, 1622 (1968).

\bibitem{DharSudarshan}
J. Dhar and E. C. G. Sudarshan,
Phys. Rev. \textbf{174}, 1808 (1968).

\bibitem{AKS}
Y. Aharonov, A. Komar, and L. Susskind,
Phys. Rev. \textbf{182}, 1400 (1969).

\bibitem{Peskin}
M. E. Peskin and D. V. Schroeder,
\textit{An Introduction to Quantum Field Theory} (Westview, Boulder, CO, 1995).

\bibitem{Sen}
A. Sen,
JHEP \textbf{9808}, 012 (1998);
JHEP \textbf{0204}, 048 (2002).

\bibitem{Tegmark}
M. Tegmark and L. Yeh,
Physica A \textbf{202}, 342 (1994).

\bibitem{Sotiriadis1}
S. Sotiriadis, P. Calabrese, and J. Cardy,
Europhys. Lett. \textbf{87}, 20 002 (2009).

\bibitem{Sotiriadis2}
S. Sotiriadis and J. Cardy,
Phys. Rev. B \textbf{81}, 134305 (2010).

\bibitem{LCpropagation}
L. Bonnes, F.H.L. Essler and A. M. L\"auchli,
Phys. Rev. Lett. \textbf{113}, 187203 (2014).

\bibitem{CardyCalabrese}
P. Calabrese and J. Cardy,
J. Stat. Mech. (2016) 064003.

\bibitem{GGE}
M. Rigol et al.,
Phys. Rev. Lett. \textbf{98}, 50405 (2007);
Nature \textbf{452}, 854 (2008).

\bibitem{Higgs}
F. Englert and R. Brout,
Phys. Rev. Lett.13, 321 (1964);
P. W. Higgs,
Phys. Rev. Lett.13, 508 (1964);
G. S. Guralnik, C. R. Hagen, and T. W. B. Kibble,
Phys. Rev. Lett.13, 585 (1964).

\bibitem{noteH1}
Note also that for a regular massive Hamiltonian $\xi\mapsto i\omega$, the unbounded term can be eliminated via a bosonic Bogoliubov transformation.

\bibitem{Sakurai}
J. J. Sakurai,
Modern Quantum Mechanics (Addison-Wesley, MA, 1994), 2nd ed.

\bibitem{Murayama}
H. Murayama,
Lecture notes on Quantum Mechanics: Path integrals, http://hitoshi.berkeley.edu/221a/.

\bibitem{Bruneton}
J.-P. Bruneton,
Phys. Rev. D \textbf{75}, 085013 (2007).

\bibitem{LRoriginal}
E. H. Lieb and D. W. Robinson,
Commun. Math. Phys. \textbf{28}, 251 (1972).

\bibitem{Nachtergaele}
B. Nachtergaele et al.,
Commun. Math. Phys. \textbf{286}, 1073 (2009);
Rev. Math. Phys. \textbf{22}, 207 (2010).

\bibitem{LRbosons}
M. Cramer, A. Serafini, and J. Eisert,
Proceedings of "Quantum Information and Many-Body Quantum Systems" Workshop, 26-31 Mar 2007, Pisa. e-print arXiv:0803.0890v2.

\bibitem{EntropyBosons}
M. Cramer, J. Eisert, M.B. Plenio, and J. Drei\ss ig,
Phys. Rev. A \textbf{73}, 012309 (2006).

\bibitem{EntropyBosonsReview}
J. Eisert, M. Cramer, and M. B. Plenio,
Rev. Mod. Phys. \textbf{82}, 277 (2010).

\bibitem{Demarie}
T. F. Demarie. e-print arXiv:1209.2748.

\bibitem{MutualInfo}
M. M. Wolf, F. Verstraete, M. B. Hastings and J. I. Cirac,
Phys. Rev. Lett.\textbf{100}, 070502 (2008).

\bibitem{Affleck}
I.Affleck,
Les Houches Lecture Notes, in: Fields, Strings, and Critical Phenomena, ed. E. Brezin and J. Zinn-Justin (North-Holland, Amsterdam, 1988).

\bibitem{phi4}
S.-J. Chang,
Phys. Rev. D \textbf{12}, 1071 (1975).

\bibitem{turbulence}
G. Sivashinsky,
Physica D \textbf{17}, 243 (1985).


\end{thebibliography}
\end{document}